\documentclass[12pt]{article}
\newcommand{\mysection}{\setcounter{equation}{0}\section}

\def\beq{\begin{equation}}
\def\eeq{\end{equation}}
\def\beqa{\begin{eqnarray}}
\def\eeqa{\end{eqnarray}}
 
\newlength{\dinwidth} \newlength{\dinmargin}
\begin{document}

\begin{center}
{\Large \bf Soft-gluon corrections in hard-scattering processes through NNNLO}
\end{center}
\vspace{2mm}
\begin{center}
{\large Nikolaos Kidonakis\footnote{Presented at the  XIV International
Workshop on Deep Inelastic Scattering (DIS 2006), Tsukuba, Japan,
April 20-24, 2006.}}\\
\vspace{2mm}
{\it Kennesaw State University, Physics \#1202 \\
1000 Chastain Rd., Kennesaw, GA 30144-5591, USA}
\end{center}

\vspace{4mm}
 
\begin{abstract}
I discuss soft-gluon corrections in hard-scattering 
processes and their resummation. I present master formulas for the 
expansion of the resummed cross section through NNNLO and discuss the  
significance of these corrections in a number of processes.
\end{abstract}
 
\thispagestyle{empty} \newpage \setcounter{page}{2}

\mysection{Introduction}

The cross section for the production of a final state $F$ in collisions 
of hadrons $h_1$ and $h_2$, 
$h_1+h_2 \rightarrow F(p) + X$, 
can be written in factorized form as
$\sigma=\sum_f \int [ \prod_i  dx_i $
\linebreak
$\phi_{f/h_i}(x_i,\mu_F)] {\hat \sigma}(s,t_i,\mu_F,\mu_R)$
where $\phi$ are the parton densities and ${\hat \sigma}$ is 
the perturbatively calculable partonic hard-scattering cross section. 
Near threshold for the production of $F$ there is
restricted phase space for real gluon emission and thus 
incomplete cancellation of infrared divergences
between real and virtual graphs resulting in the 
appearance of large logarithms in the perturbative series.
These soft and collinear logarithmic corrections take the form of
plus distributions. 
For the partonic reaction $f_1(p_1)+f_2(p_2) \rightarrow F(p) + X$
we define $s=(p_1+p_2)^2$, $t_1=(p_1-p)^2$, $t_2=(p_2-p)^2$, and 
$s_4=s+t_1+t_2-\sum m^2$. At threshold $s_4 \rightarrow 0$.
The plus didtributions are of the form   
${\cal D}_l(s_4)\equiv[\ln^l(s_4/M^2)/s_4]_+$ 
with $l \le 2n-1$ for the $n$-th order corrections in $\alpha_s$.

If we define moments of the cross section 
$\hat{\sigma}(N)=\int_0^{\infty} ds_4 e^{-Ns_4/M^2} {\hat\sigma}(s_4)$ 
then the soft corrections become logarithms of the moment variable $N$: 
${\cal D}_l(s_4)\equiv\left[\frac{\ln^{l}(s_4/M^2)}{s_4}\right]_+$
$\rightarrow \frac{(-1)^{l+1}}{l+1}\ln^{l+1}N +\cdots$. 
We can formally resum these logarithms to all orders in 
$\alpha_s$ by factorizing the soft gluons from the hard 
scattering \cite{NKres,LOS}.
To obtain physical cross sections we need to invert the moment-space 
resummed cross section back to momentum space. Resummation prescriptions 
are needed to deal with the Landau singularity.
Theoretical ambiguities are involved, and differences between prescriptions 
can be numerically bigger than higher-order terms. 

Alternatively, we can expand the resummed cross section to finite 
order \cite{NKuni}. 
No prescription is then necessary and no further approximation 
is imposed on the kinematics. 
In the expansion at next-to-leading order (NLO) in $\alpha_s$, we encounter 
${\cal D}_1(s_4)$ and ${\cal D}_0(s_4)$ terms.
At next-to-next-to-leading order (NNLO) we have ${\cal D}_3(s_4)$ 
through ${\cal D}_0(s_4)$ terms.
At next-to-next-to-next-to-leading order (NNNLO) we find 
${\cal D}_5(s_4)$ through ${\cal D}_0(s_4)$ terms.
The highest-power logarithms at each order are the leading logarithms (LL), 
the second highest are the next-to-leading logarithms (NLL), etc. 

The threshold resummation formalism has been applied by now to many processes 
including heavy quark hadroproduction \cite{NKtop}, jet production \cite{jet}, 
and electroweak processes \cite{ew}.
The numerical results invariably show that the 
soft corrections are a good approximation of the full NLO result, that 
higher-order corrections are sizable, and that the scale dependence is 
decreased dramatically when these corrections are included.

\mysection{Threshold resummation}
 
A unified formula for the resummed cross section for arbitrary processes is 
\beqa
{\hat{\sigma}}^{res}(N) &=&
\exp\left[\sum_i E^{f_i}(N_i)\right] \, 
\exp \left[\sum_i 2\int_{\mu_F}^{\sqrt{s}} \frac{d\mu}{\mu}\;
\gamma_{i/i}\left(\alpha_s(\mu)\right)\right] 
\nonumber \\ && \hspace{-20mm}\times \, 
\exp\left[\sum_j {E'}^{f_j}(N_j)\right] \;
\exp\left[2\, d_{\alpha_s} \int_{\mu_R}^{\sqrt{s}}\frac{d\mu}{\mu}\; 
\beta\left(\alpha_s(\mu)\right)\right] 
{\rm Tr} \left\{H^{f_i f_j}(\alpha_s(\mu_R)) \right.
\nonumber \\ && \hspace{-20mm}\times \, \left.
\exp \left[\int_{\sqrt s}^{\frac{\sqrt s}{{\tilde N}_j}} {d\mu \over \mu} \,
\Gamma_S^{\dagger\, f_i f_j}(\mu) \right] \,
{\tilde S}^{f_i f_j}\left(\frac{\sqrt s}{{\tilde N}_j}\right) \,
\exp \left[\int_{\sqrt s}^{\frac{\sqrt s}{{\tilde N}_j}} {d\mu \over \mu} \,
\Gamma_S^{f_i f_j}(\mu) \right] \right\}.
\eeqa
The sum over $i$ ($j$) is over incoming (outgoing) partons.
The exponents $E^{f_i}$ and ${E'}^{f_j}$ resum collinear contributions
from the incoming and outgoing partons in the hard scattering and 
are given explicitly in Ref. [3]. 
$H^{f_i f_j}$ are hard scattering matrices in color space while 
$S^{f_i f_j}$ are soft matrices that desribe noncollinear soft-gluon emission 
and whose evolution is given by the soft anomalous dimension matrices 
$\Gamma_S^{f_i f_j}$ \cite{NKres}.

\mysection{NNNLO master formulas}

Expanding the resummed cross section through NNNLO and inverting back to 
momentum space, we derive master formulas \cite{NKuni} for the soft-gluon 
corrections for arbitrary processes.

The master formula for the NLO corrections is
\beqa
{\hat{\sigma}}^{(1)} &=& \sigma^B \frac{\alpha_s(\mu_R^2)}{\pi}
\left\{c_3\, {\cal D}_1(s_4) + c_2\,  {\cal D}_0(s_4)
+c_1\,  \delta(s_4)\right\}
\nonumber \\ && 
{}+\frac{\alpha_s^{d_{\alpha_s}+1}(\mu_R^2)}{\pi}
\left[A^c \, {\cal D}_0(s_4)+T_1^c \, \delta(s_4)\right]
\eeqa
where $\sigma^B$ is the Born term,
$c_3=\sum_i 2 \, C_i -\sum_j C_j$,  
with $C_q=C_F$ for quarks and $C_g=C_A$ for gluons,
and $c_2$ is defined by $c_2=c_2^{\mu}+T_2$, 
with
$c_2^{\mu}=-\sum_i C_i \ln(\mu_F^2/M^2)$
denoting the terms involving logarithms of the scale, and  
$T_2=- \sum_i \left[C_i
+2 \, C_i \, \ln\left(\frac{-t_i}{M^2}\right)\right.$
\linebreak
$\left. +C_i \ln\left(\frac{M^2}{s}\right)\right]
-\sum_j \left[B_j^{(1)}+C_j
+C_j \, \ln\left(\frac{M^2}{s}\right)\right]$  
denoting the scale-independent terms.
Here $\mu_F$ ($\mu_R$) is the factorization (renormalization) scale, 
$M$ is a hard scale relevant to the process under study, and 
$B_j^{(1)}$ equals $3C_F/4$ for quarks and $\beta_0/4$ for gluons.
The function $A^c$ is process-dependent and depends on the color 
structure of the hard-scattering. It is defined by
$A^c={\rm tr} \left(H^{(0)} \Gamma_S^{(1)\,\dagger} S^{(0)}
+H^{(0)} S^{(0)} \Gamma_S^{(1)}\right)$.
With regard to the $\delta(s_4)$ terms, we split them into a term
$c_1$, that is proportional 
to the Born cross section, and a term $T_1^c$ that is not.

The master formula for the NNLO corrections is 
\beqa
{\hat{\sigma}}^{(2)}&=&
\sigma^B \frac{\alpha_s^2}{\pi^2} 
\frac{1}{2} c_3^2 {\cal D}_3(s_4)
\nonumber \\ && \hspace{-12mm}
{}+\sigma^B \frac{\alpha_s^2}{\pi^2} 
\left\{\frac{3}{2} c_3  c_2 - \frac{\beta_0}{4} c_3
+\sum_j C_j  \frac{\beta_0}{8}\right\}  {\cal D}_2(s_4)
+\frac{\alpha_s^{d_{\alpha_s}+2}}{\pi^2} 
\frac{3}{2}  c_3  A^c {\cal D}_2(s_4)
\nonumber \\ && \hspace{-12mm}
{}+\sigma^B \frac{\alpha_s^2}{\pi^2} C_{D_1}^{(2)}
{\cal D}_1(s_4)+\frac{\alpha_s^{d_{\alpha_s}+2}}{\pi^2}
\left\{\left(2 c_2-\frac{\beta_0}{2}\right) A^c+c_3  T_1^c
+F^c\right\} {\cal D}_1(s_4)
\nonumber \\ && \hspace{-12mm}
{}+\cdots
\eeqa
where 
\beq
C_{D_1}^{(2)}=c_3 \, c_1 +c_2^2
-\zeta_2 \, c_3^2 -\frac{\beta_0}{2} \, T_2 
+\frac{\beta_0}{4} \, c_3 \, \ln\left(\frac{\mu_R^2}{M^2}\right)
+c_3\, \frac{K}{2} 
-\sum_j\frac{\beta_0}{4} \, B_j^{(1)} \, , 
\eeq
with $K=C_A(67/18-\pi^2/6)-5n_f/9$, and 
\beq
F^c={\rm tr} \left[H^{(0)} \left(\Gamma_S^{(1)\,\dagger}\right)^2 S^{(0)}
+H^{(0)} S^{(0)} \left(\Gamma_S^{(1)}\right)^2
+2 H^{(0)} \Gamma_S^{(1)\,\dagger} S^{(0)} \Gamma_S^{(1)} \right] \, .
\label{Fterm}
\eeq

The master formula for the NNNLO corrections is 
\beqa
{\hat{\sigma}}^{(3)}&=&
\sigma^B \frac{\alpha_s^3}{\pi^3} 
\frac{1}{8} c_3^3 {\cal D}_5(s_4)
\nonumber \\ && \hspace{-12mm}
{}+\sigma^B \frac{\alpha_s^3}{\pi^3} 
\left\{\frac{5}{8} c_3^2 c_2 -\frac{5}{2} c_3 X_3\right\} 
{\cal D}_4(s_4)
+\frac{\alpha_s^{d_{\alpha_s}+3}}{\pi^3} 
\frac{5}{8}  c_3^2  A^c  {\cal D}_4(s_4)
\nonumber \\ && \hspace{-12mm}
{}+\sigma^B \frac{\alpha_s^3}{\pi^3}
\left\{c_3 c_2^2 +\frac{c_3^2}{2} c_1
-\zeta_2 c_3^3 +(\beta_0-4c_2) X_3 +2 c_3 X_2
-\sum_j C_j \frac{\beta_0^2}{48}\right\} {\cal D}_3(s_4)
\nonumber \\ && \hspace{-12mm}
{}+\frac{\alpha_s^{d_{\alpha_s}+3}}{\pi^3}
\left\{\frac{1}{2} c_3^2 T_1^c+\left[2 c_3 c_2
-\frac{\beta_0}{2} c_3 -4  X_3 \right] A^c +c_3 F^c\right\}
{\cal D}_3(s_4)
+ \cdots
\eeqa
where 
$X_3=\beta_0 c_3/12-\sum_j C_j \beta_0/24$ and  
$X_2=-(\beta_0/4)T_2+(\beta_0/8)c_3 \ln(\mu_R^2/M^2)
+c_3 K/4-\sum_j \beta_0 B_j^{(1)}/8$.

The formalism has been applied recently to top quark production at the 
Tevatron \cite{NKuni,NKtop}. The corrections are non-negligible and serve 
to substantially reduce the scale dependence of the cross section. 
The theoretical results are in excellent agreement with data from 
the CDF \cite{CDF} and D0 \cite{D0} experiments.

\end{document}